\documentstyle[12pt,psfig,amsfonts,amssymb]{elsart}

\begin{document}
\begin{frontmatter}

\title{Covariant nucleon electromagnetic form factors from the
Goldstone-boson-exchange quark model}

\author[Graz]{R.F.~Wagenbrunn}, 
\author[Pavia]{S.~Boffi}, 
\author[Iowa]{W.~Klink},
\author[Graz]{W.~Plessas} and
\author[Pavia]{M.~Radici} 

\address[Graz]{Institut f\"ur Theoretische Physik, Universit\"at Graz, \\
Universit\"atsplatz 5, A-8010 Graz, Austria}

\address[Pavia]{Dipartimento di Fisica Nucleare e Teorica,
Universit\`a di Pavia \\ and Istituto Nazionale di Fisica Nucleare, 
Sezione di Pavia, I-27100 Pavia, Italy}
\address[Iowa]{Department of Physics and Astronomy, University of Iowa, \\
Iowa City, IA 52242, USA}

\begin{abstract}
We present a study of proton and neutron electromagnetic form factors for the
recently proposed Goldstone-boson-exchange constituent quark model. Results for
charge radii, magnetic moments, and electric as well as magnetic form factors are
reported. The calculations are performed in a covariant framework using
the point-form approach to relativistic quantum mechanics. All the predictions by
the Goldstone-boson-exchange constituent quark model are found in remarkably good
agreement with existing experimental data.

\vspace{.2cm}
\noindent
{\sl PACS\/}: 12.39-x; 13.40.-f; 14.20.Dh

\noindent
{\sl Keywords\/}: Nucleon electromagnetic form factors; Constituent quark model;
                  Point-form quantum mechanics
\end{abstract}

\end{frontmatter}

\section{Introduction and motivation}
\label{intro}
Constituent quark models (CQM) provide a promising tool for the description of
low-energy hadron phenomena. They permit the introduction of essential properties
of nonperturbative quantum chromodynamics (QCD) and provide a 
framework for quantitative calculations of
hadron properties and reaction observables. In particular, CQM can be made to 
incorporate the basic symmetries of QCD and, at the same time, of relativistic
covariance.

CQM a priori aim at an effective description of QCD rather than its direct
solution. They rely on the assumption of constituent quarks ($Q$), which are
viewed as quasiparticles generated dynamically by the spontaneous breaking of
chiral symmetry (SB$\chi$S). This understanding is in line with the general
ideas of dynamical symmetry breaking (e.g., following the ideas of Nambu and
Jona-Lasinio \cite{Nambu:1961tp}) and has recently been
further substantiated by lattice QCD
results \cite{Aoki:1999mr}.

Over the years there has been an ongoing search for the features of the (effective)
$Q$-$Q$ interaction, specifically in nucleons and more generally in all light as well
as strange baryons and their resonances. The debate has been intensified by the
appearance of the CQM based on Goldstone-boson-exchange (GBE) dynamics
\cite{Glozman:1998ag}. This type of CQM assumes a linear confinement, as suggested
by lattice QCD, with a strength according to the string tension of QCD. For the
hyperfine interaction it exploits the idea of GBE \cite{Glozman:1996fu}. Due to 
SB$\chi$S, the original $\rm SU(3)_L\times SU(3)_R$ symmetry of QCD is reduced to
an $\rm SU(3)_V$ (vector) symmetry associated with the appearance of Goldstone bosons.
As a result constituent quarks and Goldstone bosons form the essential degrees of
freedom governing the low-energy physics of light and strange baryons.

The version of the GBE CQM in Ref. \cite{Glozman:1998ag} uses a
relativistic kinetic-energy operator and advocates, in addition to the linear
confinement potential, a pseudoscalar GBE hyperfine interaction;
for the latter only the spin-spin component is taken,
which is most important in the hyperfine splitting of
the baryon spectra. Due to the
specific spin-flavor dependence of the hyperfine interaction this kind of CQM has
been remarkably successful in reproducing the detailed features of the excitation
spectra of all light and strange baryons. Within a single model it has been
possible to describe all the resonance levels with the correct orderings.
There is no need to separate positive- and negative-parity excitations into
different parametrizations, and the particular flavor dependence allows
the different level orderings in the $N$ and $\mit\Lambda$ spectra to be described
simultaneously.

Baryon spectroscopy, however, is only a first, though quite demanding, test of
low-energy models of QCD. Furthermore, CQM should also provide for a
comprehensive description of other hadron phenomena, such as
electromagnetic (e.m.) nucleon form factors, resonance excitations and decays,
etc. With regard to the GBE
CQM the question is whether it can yield as successful a description of
these reactions as was the case for spectroscopy. Here we give an answer
concerning nucleon e.m. form factors.

In this paper we report first results of proton and neutron electric as well as
magnetic form
factors calculated with the GBE CQM wave functions in point-form relativistic
quantum mechanics. Due to the fact that the GBE CQM uses a relativistic
kinetic-energy operator it also provides a covariant mass operator containing
interactions given by a Bakamjian--Thomas (BT)
construction \cite{Bakamjian:1953kh}. Even though the
total $Q$-$Q$ interaction consists of a phenomenological confinement and an
instantaneous one-boson-exchange potential, both essentially
nonrelativistic, the full Hamiltonian leads to a mass operator fulfilling
all necessary commutation relations of the Poincar\'e group \cite{Keister:1991sb}.
From the various possibilities for setting up a relativistic quantum theory
\cite{Dirac:1949cp}, we have chosen the point-form formulation.
It is characterized by several distinctive features. All the dynamics is contained
in the four-momentum operators, which commute among themselves and can be
simultaneously diagonalized. The generators of the Lorentz boosts contain no
interactions and so are purely kinematic; the theory is thus manifestly
covariant. In practice, the point-form approach allows us to properly
perform Lorentz boosts of the three-$Q$ wave functions and to accurately
calculate the matrix elements of the e.m. current operator. In particular, by the
introduction of so-called velocity states \cite{Klink:1998hc} we can carry out all
necessary transformations of the momentum dependences in the wave functions and
the relativistic quark spin rotations associated with boosting the nucleon state.

Here we follow the formalism developed in Ref. \cite{Klink:1998hb} to investigate
the e.m. structure of the nucleons. We calculate the proton and neutron
e.m. form factors in point-form relativistic quantum mechanics with
one-body currents only. This approach corresponds to a relativistic impulse
approximation but specifically in point form. It is called point-form
spectator approximation (PFSA) \cite{Allen:2000ge} to distinguish it from
impulse approximations in other forms of relativistic quantum mechanics,
which may lead to different results (see the discussion in Section~\ref{results}). 

In the following we briefly outline the calculations of e.m. current matrix
elements with three-$Q$ wave functions in the point-form formulation. The
corresponding results for proton as well as neutron electric and magnetic Sachs
form factors, charge radii, and magnetic moments are presented in Section
\ref{results}. A summary and our conclusions are given in Section \ref{sac}.

\section{Nucleon form factors in point form}
\label{pf}
We start with the eigenstates of the quark-model Hamiltonian for baryons
\begin{equation}
\label{eq:h}
H=\sum\limits_{i=1}^3\sqrt{\vec{k}_i^2+m_i^2}+\sum\limits_{i<j=1}^3
\left[V^{\rm conf}(i,j)+V^{\rm hf}(i,j)\right].
\end{equation}
Herein the first term, with $m_i$ the masses and $\vec{k}_i$ the three-momenta of
the constituent quarks, represents the relativistic kinetic energy. The $Q$-$Q$
interaction consists of the confinement and hyperfine potentials. Specifically we
adhere to the version of the GBE CQM as presented in Ref.~\cite{Glozman:1998ag}.

The three-$Q$ Hamiltonian (\ref{eq:h}) is solved using the stochastic variational
method (SVM) \cite{Suzuki:1998aa}. This approach yields the eigenenergies with
great accuracy; the corresponding results for the ground-state and resonance
energies have recently been confirmed by a completely independent method, namely
by solving modified Faddeev equations \cite{Papp:2000kp}. Moreover, the SVM also
produces the three-$Q$ eigenstates in the center-of-momentum frame, i.e.
for total nucleon three-momentum $\vec{P}=0$. The corresponding wave functions
can then be interpreted as eigenstates of the mass operator including interactions
\cite{Klink:1998hb}
\begin{equation}
\label{eq:mass}
M=\sqrt{P^\mu P_\mu}=M_{\rm free}+M_{\rm int}.
\end{equation}
Here, $P^\mu=P^\mu_{\rm free}+P^\mu_{\rm int}$ is the four-momentum
operator with interactions according to the BT construction in point form.
We write the general three-$Q$ state defined on the product space
${\mathcal H}_1\otimes{\mathcal H}_2\otimes{\mathcal H}_3$
of one-particle spin-$1\over 2$, positive-mass, positive-energy representations
${\mathcal H}_i=L^2({\mathbb R}^3)\times S^{1/2}$ of the Poincar\'e group as
\begin{equation}
\label{eq:prodstate}
\left|p_1,p_2,p_3;\lambda_1,\lambda_2,\lambda_3\right\rangle=
\left|p_1,\lambda_1\right\rangle\otimes
\left|p_2,\lambda_2\right\rangle\otimes
\left|p_3,\lambda_3\right\rangle,
\end{equation}
where $p_i$ are the individual quark four-momenta and $\lambda_i$ the
$z$-projections of their spins. Under general Lorentz transformations
$U_\Lambda$ these states transform as
\begin{equation}
\label{eq:boost1}
U_\Lambda \left|p_1,p_2,p_3;\lambda_1,\lambda_2,\lambda_3\right\rangle=
\prod\limits_{i=1}^{3}D^{1/2}_{\lambda'_i\lambda_i}(R_{W_i})
\left|\Lambda p_1,\Lambda p_2,\Lambda
p_3;\lambda'_1,\lambda'_2,\lambda'_3\right\rangle.
\end{equation}
In this equation sums over all $\lambda'_1,\lambda'_2,\lambda'_3$
are understood (here and in the following we adhere to the usual convention
of summing over identical indices). The Lorentz transformation in
Eq. (\ref{eq:boost1}) involves three different Wigner rotations.
It is more convenient to first introduce so-called velocity states
\cite{Klink:1998hc} by applying a particular Lorentz boost $U_{B(v)}$ to the
center-of-momentum states, which are defined analogously to
Eq. (\ref{eq:prodstate}) but fulfil the constraint 
$\vec{P}=\vec{k_1}+\vec{k_2}+\vec{k_3}=0$,
\begin{equation}
\label{eq:vstate}
\left|v;\vec{k}_1,\vec{k}_2,\vec{k}_3;\mu_1,\mu_2,\mu_3\right\rangle=U_{B(v)}
\left|k_1,k_2,k_3;\mu_1,\mu_2,\mu_3\right\rangle.
\end{equation}
Under general Lorentz transformations $U_\Lambda$ these velocity states
transform as
\begin{eqnarray}
\label{eq:boost2}
&&U_\Lambda\left|v;\vec{k}_1,\vec{k}_2,\vec{k}_3;\mu_1,\mu_2,\mu_3\right\rangle
=
U_\Lambda U_{B(v)}\left|k_1,k_2,k_3;\mu_1,\mu_2,\mu_3\right\rangle=\nonumber\\
&=&
U_{B(\Lambda v)}U_{R_W}\left|k_1,k_2,k_3;\mu_1,\mu_2,\mu_3\right\rangle
=\nonumber\\
&=&
\prod\limits_{i=1}^3D^{1/2}_{\mu'_i\mu_i}[R_W(k_i,R_W)]
\left|\Lambda v;R_W\vec{k}_1,R_W\vec{k}_2,R_W\vec{k}_3;\mu'_1,\mu'_2,\mu'_3\right\rangle,
\end{eqnarray}
where $R_W$ is the Wigner rotation $R_W(v,\Lambda)$ and $R_W(k_i,R_W)$
is the Wigner rotation of a Wigner rotation. If the boost $B(k_i)$ is chosen to be
a canonical one, then $R_W(k_i,R_W)=R_W$ \cite{Klink:1992ui}, and one has
\begin{eqnarray}
\label{eq:boost3}
&&U_\Lambda\left|v;\vec{k}_1,\vec{k}_2,\vec{k}_3;\mu_1,\mu_2,\mu_3\right\rangle=
\nonumber\\
&=&\prod\limits_{i=1}^3D^{1/2}_{\mu'_i\mu_i}(R_W)
\left|\Lambda v;R_W\vec{k}_1,R_W\vec{k}_2,R_W\vec{k}_3;\mu'_1,\mu'_2,\mu'_3\right\rangle.
\end{eqnarray}
Here the Wigner rotations are all the same and the spins can thus be coupled
together to a total spin state as in nonrelativistic theory. If a boost other than
canonical spin is applied, one can modify Eq.~(\ref{eq:vstate}) and still arrive at
the same transformation property as in Eq.~(\ref{eq:boost3}).
The connection between velocity states and the general three-$Q$ states of
Eq.~(\ref{eq:prodstate}) is given by
\begin{equation}
\label{eq:v3q}
\left|v;\vec{k}_1,\vec{k}_2,\vec{k}_3;\mu_1,\mu_2,\mu_3\right\rangle=
\prod\limits_{i=1}^3D^{1/2}_{\lambda_i\mu_i}[R_W(k_i,B(v))]
\left|p_1,p_2,p_3;\lambda_1,\lambda_2,\lambda_3\right\rangle
\end{equation}
with $p_i=B(v)k_i$.

In Eqs.~(\ref{eq:vstate})--(\ref{eq:v3q}) one may equally well express the
velocity states in terms of momenta $\vec{p}$ and $\vec{q}$ conjugate to the Jacobi
coordinates $\vec{x}$ and $\vec{y}$ (of a certain three-particle partition).
The Lorentz transformation in Eq.~(\ref{eq:boost3}) then specifically reads as
\begin{equation}
\label{eq:boost4}
U_\Lambda\left|v;\vec{p},\vec{q};\mu_1,\mu_2,\mu_3\right\rangle=
\prod\limits_{i=1}^3D^{1/2}_{\mu'_i\mu_i}(R_W)
\left|\Lambda v;R_W\vec{p},R_W\vec{q};\mu'_1,\mu'_2,\mu'_3\right\rangle.
\end{equation}

Once the Lorentz transformations on the three-$Q$ states have been dealt with,
the next task is to calculate the matrix elements of the electromagnetic current
operator. For this purpose we follow the formalism developed in
Ref.~\cite{Klink:1998hb}. We first write the invariant nucleon form factors as
matrix elements of the one-particle current operator $j^\nu_{[1]}$
(i.e. in the PFSA) in the standard Breit frame, where the momentum 
transfer $Q$ on the nucleon is only in $z$-direction $(0,0,0,Q)=q_{\rm st}$, as
\begin{eqnarray}
\label{eq:invff}
F_{\mu'\mu}^\nu(Q^2)&=&3\int
d^3pd^3qd^3p'd^3q'\psi^\ast_{\mu'}(\vec{p}\,',\vec{q}\,';\mu'_1,\mu'_2,\mu'_3)
\psi_{\mu}(\vec{p},\vec{q};\mu_1,\mu_2,\mu_3)\nonumber\\
&&\hspace{-2em}\times{D^{1/2}_{\lambda'_1\mu'_1}}^\ast[R_W(k'_1,B(v_{\rm out}))]
\left\langle p'_1,\lambda'_1\right|j^\nu_{[1]}\left|p_1,\lambda_1\right\rangle
D^{1/2}_{\lambda_1\mu_1}[R_W(k_1,B(v_{\rm in}))]\nonumber\\
&&\hspace{-2em}\times D^{1/2}_{\mu'_2\mu_2}[R_W(k_2,B^{-1}(v_{\rm out})B(v_{\rm in}))]
D^{1/2}_{\mu'_3\mu_3}[R_W(k_3,B^{-1}(v_{\rm out})B(v_{\rm in}))]\nonumber\\
&&\hspace{-2em}\times\delta^3[k'_2-B^{-1}(v_{\rm out})B(v_{\rm in})k_2]
\delta^3[k'_3-B^{-1}(v_{\rm out})B(v_{\rm in})k_3].
\end{eqnarray}
Due to the symmetry of the wave functions it is sufficient to
consider only the case where quark 1 is struck by the photon, while
quarks 2 and 3 are the spectators, and to multiply the result by 3.
The initial and final velocities are given by
$m_N v_{\rm in}=p_{\rm st}=(\sqrt{m_N^2+(Q/2)^2},0,0,-Q/2)$ and
$m_N v_{\rm out}=p'_{\rm st}=(\sqrt{m_N^2+(Q/2)^2},0,0,Q/2)$, respectively,
where $m_N$ denotes the nucleon mass.

The single-particle current matrix element in Eq.~(\ref{eq:invff})
has the usual form \cite{Klink:1998hb}
\begin{equation}
\label{eq:current}
\left\langle p'_i,\lambda'_i\right|j^\nu_{[1]}\left|p_i,\lambda_i\right\rangle=
e_1\bar{u}(p'_i,\lambda'_i)
\left[\gamma^\nu f_1(\tilde{Q}^2)+i
\frac{\sigma^{\nu\rho}\tilde{q}_\rho}{2m_i}
f_2(\tilde{Q}^2)\right]u(p_i,\lambda_i)
\end{equation}
with $u(p_i,\lambda_i)$ the Dirac spinor of quark $i$ and
$\tilde{q}_\rho=p'_\rho-p_\rho$, $\tilde{Q}^2=-\tilde{q}^2$, the 
momentum transfer on a single quark.
It contains the quark invariant form factors $f_1$
and $f_2$. In the present study we assume pointlike constituent quarks for which
$f_1(\tilde{Q}^2)=1$ and  $f_2(\tilde{Q}^2)=0$.

From Eq.~(\ref{eq:invff}) one obtains the nucleon Sachs form factors through
\begin{equation}
\label{eq:sachs}
\left.\begin{array}{l}
\displaystyle F^{\nu=0}_{\mu'\mu}(Q^2)=G_E(Q^2)\delta_{\mu',\mu}\\
\displaystyle F^{\nu=2}_{\mu'\mu}(Q^2)=\frac{Q}{m_N}G_M(Q^2)\delta_{\mu',\mu\pm 1}
\end{array}\right.\qquad
\mu,\mu'=\pm\frac{1}{2}.
\end{equation}
Note that only the $\nu=0$ and $\nu=2$ components of $F^{\nu}_{\mu'\mu}$
are needed for the electric and magnetic form factors, respectively.
There is no new information in $F^{\nu=1}_{\mu'\mu}$, and one
simply recovers $G_M$.
The $\nu=3$ component of the current vanishes
due to current conservation $q_{st}^{\nu}j_{\nu}=0$ (i.e. the 
continuity equation in the standard Breit frame).
\vspace*{-0.3cm}

\section{Results}
\label{results}
\vspace*{-0.2cm}
The predictions of the GBE CQM \cite{Glozman:1998ag} for the nucleon 
e.m. form factors
are shown in Fig.~\ref{fig:ff} below. Their properties at zero momentum
transfer are reflected by the charge radii and magnetic moments given in
Table~\ref{tab:crmm}. The results
were calculated in PFSA as explained in the previous section. The input
into the calculations consists
only of the proton and neutron three-$Q$ wave functions as produced by
the solution of the Hamiltonian (\ref{eq:h}).

\begin{figure}[t]
\vspace*{-0.3cm}
$
\begin{array}{lr}
\hspace*{-0.7cm}\psfig{file=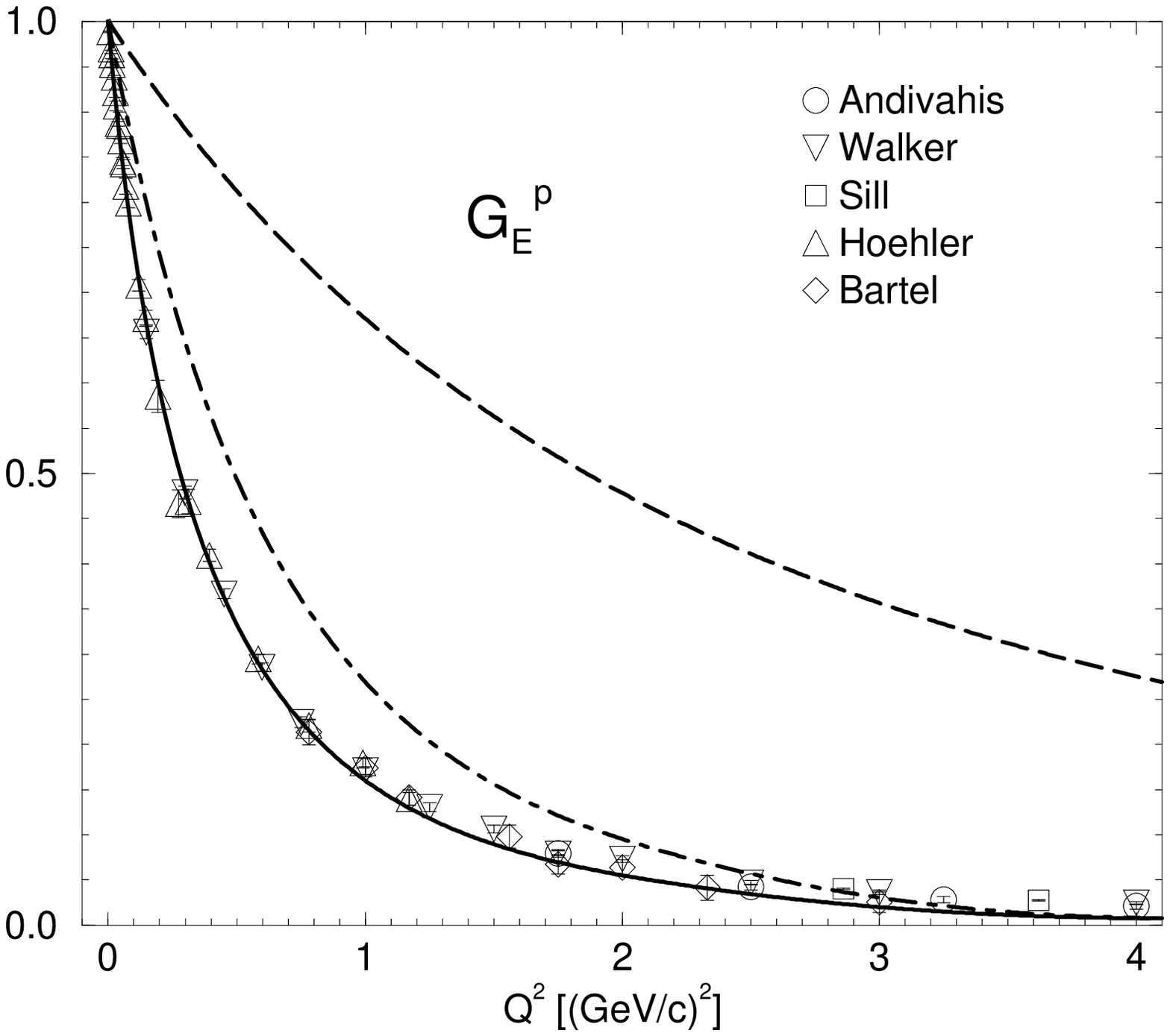,width=8.2cm}&
\hspace*{-1.3cm}\psfig{file=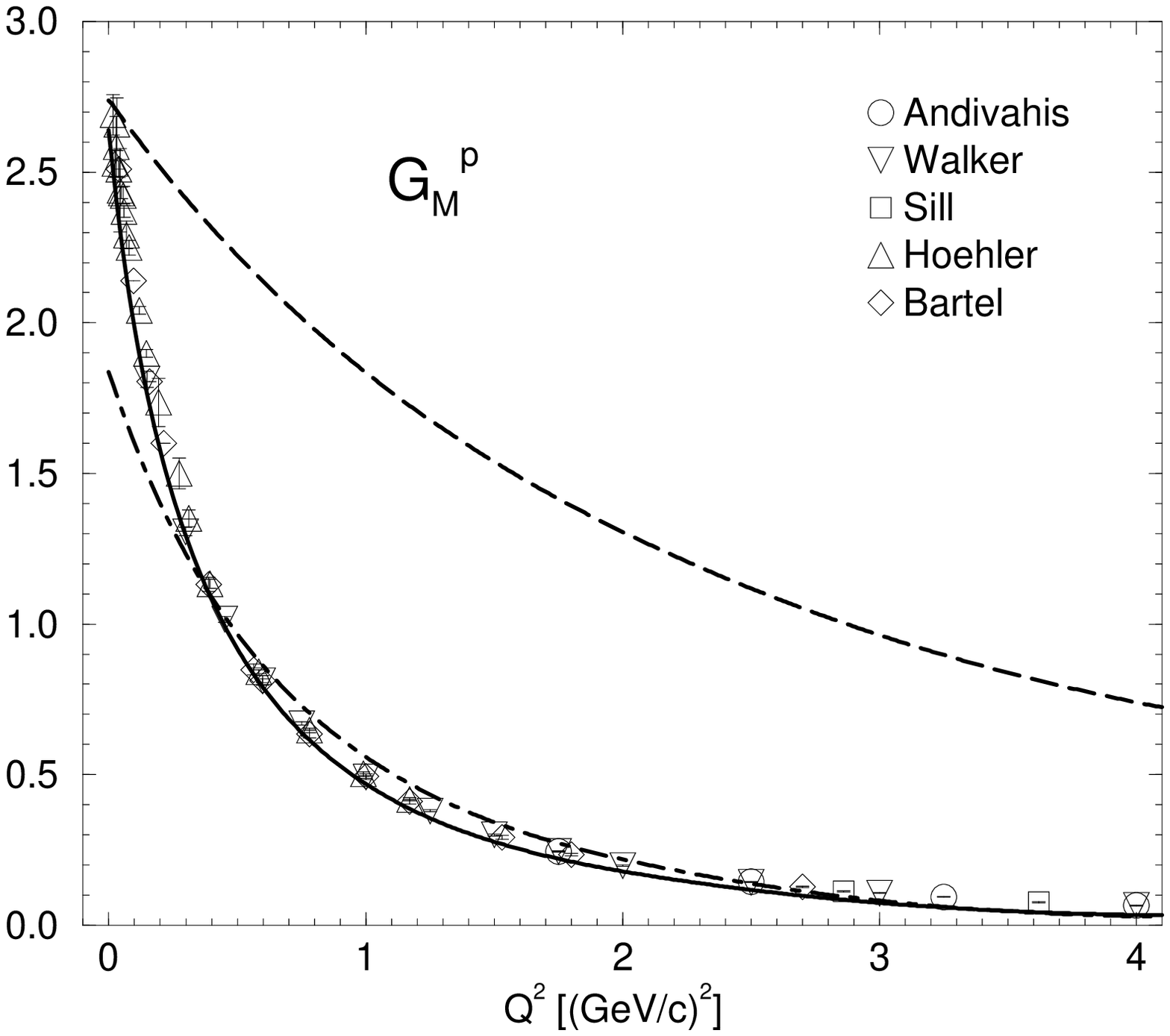,width=8.2cm}\\[-0.9cm]
\hspace*{-0.7cm}\psfig{file=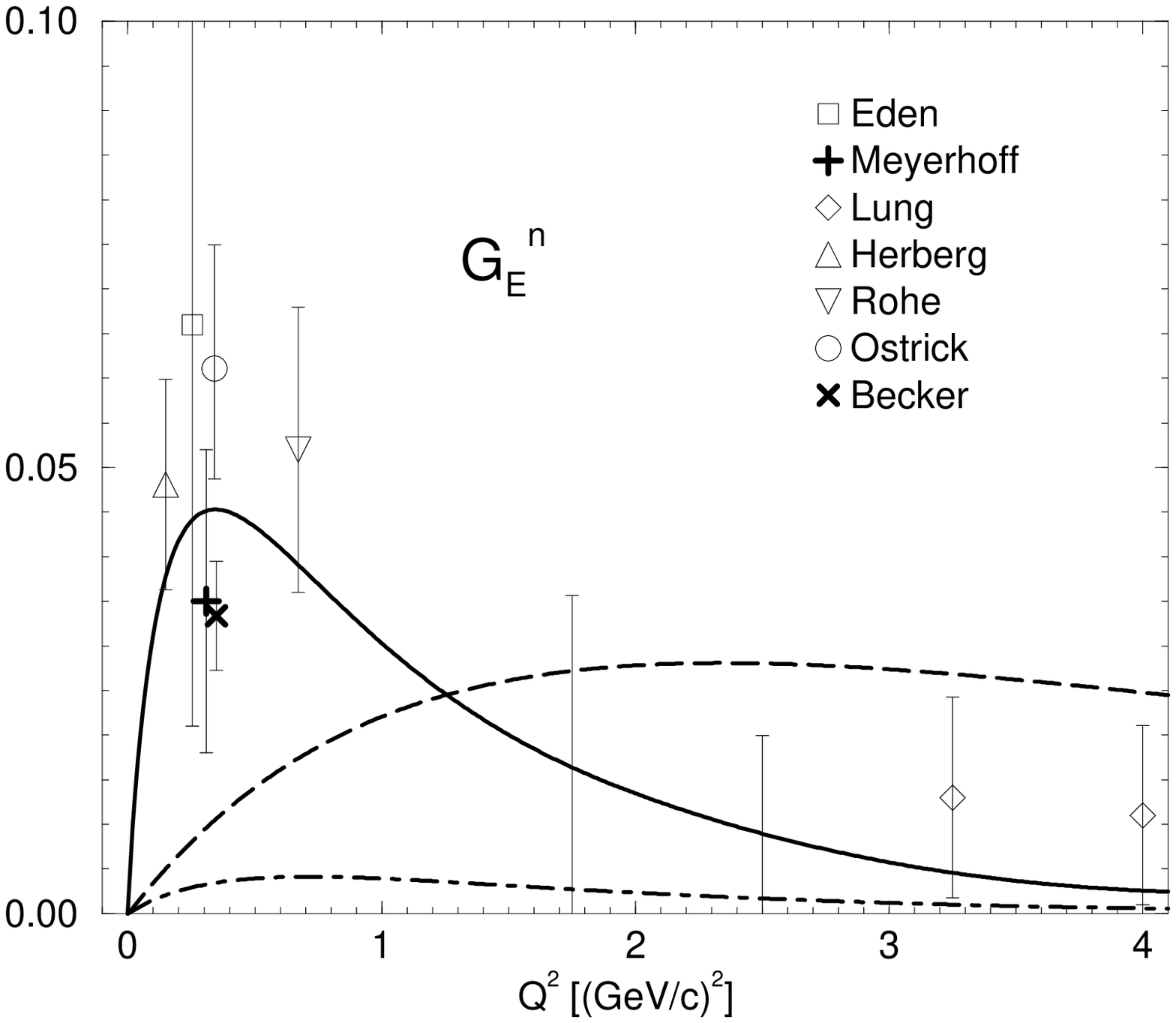,width=8.2cm}&
\hspace*{-1.3cm}\psfig{file=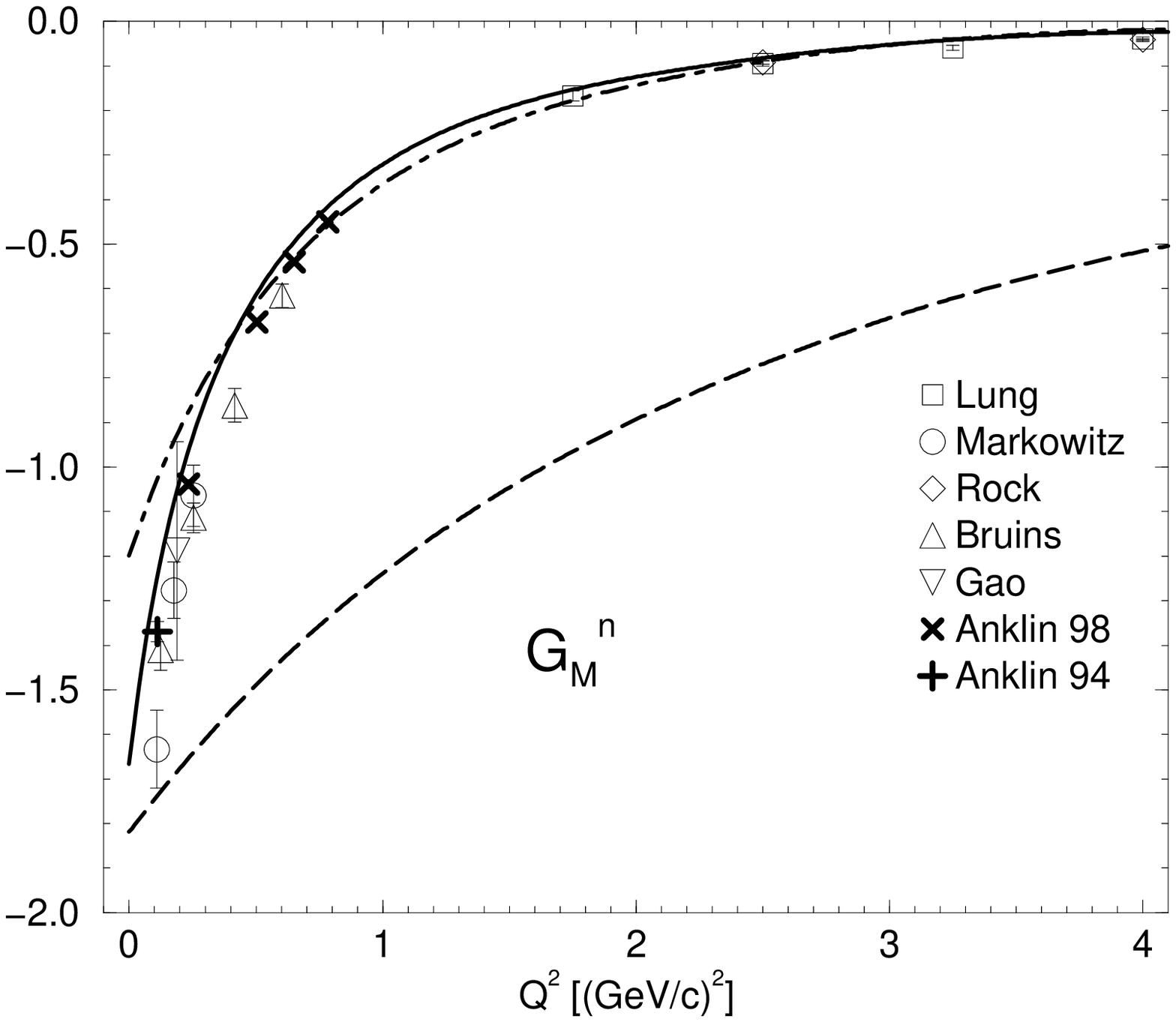,width=8.2cm}\\[-0.4cm]
\end{array}
$
\caption{Proton (upper) and neutron (lower) electric (left) and magnetic (right)
form factors as predicted by the GBE CQM \cite{Glozman:1998ag} in PFSA
(solid lines). A comparison is given to
the results in NRIA (dashed) and the case with the confinement interaction only
(dashed-dotted). The experimental data are from
Ref.~\cite{Andivahis:1994rq}.}
\label{fig:ff}
\end{figure}
\begin{table}[b]
\caption{Proton and neutron charge radii as well as magnetic moments as predicted
by the GBE CQM \cite{Glozman:1998ag} in PFSA. A comparison is given also to the
results in NRIA and the case with the confinement interaction only.}
\label{tab:crmm}
\begin{center}
\begin{tabular*}{\textwidth}{l@{\qquad}r@{\qquad}r@{\qquad}r@{\extracolsep\fill}l}
\hline
                  &  PFSA    &  NRIA    &  Conf.   & Experimental\\ \hline
$r^2_p$ [fm$^2$]  & $ 0.75$ & $ 0.10$ & $ 0.37$ & $0.774(27)$
                                                    \cite{Rosenfelder:1999cd},
						    $0.780(25)$
						    \cite{Melnikov:1999xp}\\
$r^2_n$ [fm$^2$]  & $-0.12$ & $-0.01$ & $-0.01$ & $-0.113(7)$
                                                    \cite{Kopecky:1995xp}\\
$\mu_p$ [n.m.]    & $ 2.64$ & $ 2.74$ & $ 1.84$ & $2.792847337(29) $
                                                    \cite{Groom:2000xp}\\
$\mu_n$ [n.m.]    & $-1.67$ & $-1.82$ & $-1.20$ & $-1.91304270(5)$
                                                    \cite{Groom:2000xp}\\
\hline
\end{tabular*}
\end{center}
\end{table}
One observes that an extremely good description of both the proton and
neutron e.m. structure is achieved.
It is rather surprising that all relevant observables are
quite correctly reproduced. This appears 
remarkable in view of the numerous attempts that have been made to explain the
low-momentum-transfer e.m. form factors from CQM. Evidently relativity plays a
major role here. For comparison we also show results for the form factors when
calculated in nonrelativistic impulse approximation (NRIA), i.e. with 
the standard nonrelativistic form of the current operator and no 
Lorentz boosts applied to the nucleon wave functions. Evidently there
is no way of describing the nucleon e.m. form factors in a nonrelativistic theory.
However, also in comparison to other relativistic attempts (see, e.g.,
\cite{Cardarelli:1995dc,Szczepaniak:1995mi,Coester:1997ih})
the covariant results obtained here 
with the use of realistic CQM wave functions in 
the point-form approach appear noteworthy, since the e.m. form 
factors of both the neutron and the proton are readily explained even 
with pointlike constituent quarks. At least for the range of momentum 
transfers considered in Fig.~\ref{fig:ff} there is no need 
to introduce constituent quark form factors (or any other 
phenomenological parameters beyond the CQM). This has been necessary 
in previous relativistic studies in order to bring the theoretical 
predictions in agreement with experimental data
\cite{Cardarelli:1995dc,Szczepaniak:1995mi,Coester:1997ih}.
Here, in particular, the momentum 
dependences already have the right behaviour. E.g., the proton 
electric form factor nicely matches the dipole form for 
$Q^2\lesssim 1$ GeV$^2$, while it starts to deviate from it beyond, 
following the trend of recent JLab data \cite{Jones:1999rz}. It is only 
with regard to the magnetic moments that there remains a small difference 
between the theoretical predictions and the experimental data. An 
explanation for this gap might be offered by a recent study of
pion-loop corrections to magnetic moments \cite{Glozman:1998yh}.

The results presented here depend predominantly on the relativistic 
boost effects introduced into the nucleon wave functions. The 
corresponding Lorentz transformations affect the quark spins and the 
momentum dependences of the wave functions (cf. 
Eq.~(\ref{eq:invff})). In the point-form approach we are able to 
perform these transformations without any approximations. The 
calculations are facilitated by the fact that in point form all 
interactions are contained in the momentum operators while the 
generators of the Lorentz group remain interaction-free. This represents 
an important technical advantage, in that the angular momenta and Lorentz 
boosts are just the same as in the free case.

In order to get an idea of the role of the GBE hyperfine interaction in the e.m.
form factors, we have also considered the case with the confinement potential only.
In addition to differences in the wave functions, the nucleons now 
also have a larger mass of $m_N^{\rm conf}=1353$ MeV. This different
mass is very important for the behavior of the form factors for low $Q^2$ 
and is essentially responsible for the corresponding results given in 
Table~\ref{tab:crmm}. Shifting the nucleon mass artificially to 
$m_N=939$ MeV would change the charge radii and magnetic moments
in the following way: $r^2\to r^2\left(m_N^{\rm conf}/m_N\right)^2$ and
$\mu\to\mu\left(m_N^{\rm conf}/m_N\right)$. As a result 
the proton charge radius as well as the magnetic moments of both the
proton and the neutron would then already
be very close to the values obtained with the full interaction.
Only the neutron charge radius would still remain much too small, due 
to the absence of the mixed-symmetry component in the wave function
for the case with the confinement potential only. Though the 
mixed-symmetry component brought about by the hyperfine interaction
is rather small, it turns out to be most essential for reproducing the
neutron charge radius in a reasonable manner.

\section{Summary and Conclusions}
\label{sac}
We have presented a first study of elastic nucleon e.m. form factors with the GBE
CQM in the point-form approach. The theoretical predictions obtained in PFSA are
found to be in remarkably good agreement with all experimental data (charge
radii, magnetic moments, electric and magnetic form factors) both for the proton
and the neutron. No further ingredients beyond the quark model wave functions (such
as constituent quark form factors etc.) have been employed. Only relativistic boost
effects are properly included in point-form relativistic quantum mechanics.

Our results suggest that relativistic boost effects are most important in the
calculation of nucleon e.m. observables. The point-form approach appears
advantageous from a practical point of view, as it makes it possible to 
include all
boost effects in the evaluation of the current-operator matrix elements.

After the successful description of the spectroscopy of all light and strange
baryons in a unified framework \cite{Glozman:1998ag}, the GBE CQM now appears
capable of also explaining the first dynamical observables, namely the nucleon e.m.
structure. It will be important to perform a series of further detailed studies
related to the present investigation of proton and neutron elastic form factors. At
the same time one is immediately tempted to ask how well electromagnetic
transitions (and further on hadronic reactions such as baryon resonance decays) can
be described. The point-form approach offers the possibility for performing
the relevant investigations on a reliable basis. 

\vspace{-0.3cm}
\ack{We have profited from numerous discussions with L.Ya. Glozman. The work
was partly performed under the contract ERB
FMRX-CT-96-0008 within the frame of the Training and Mobility of
Researchers Programme of the European Union.}

\end{document}